\documentclass[11pt,preprint,tightenlines,aps,showpacs,letterpaper]{revtex4}

\usepackage{graphicx}
\usepackage{slashed}

\usepackage{comment}
\usepackage{amsmath}
\usepackage{amssymb}
\usepackage[hidelinks]{hyperref}
\usepackage{tikz}
\usepackage{tikz-feynman}
\usetikzlibrary{arrows.meta}
\tikzset{
  myarrow/.style={
    decoration={markings, mark=at position 0.63 with {\arrow{Latex[length=2mm, width=1mm]}}},
    postaction={decorate}
  }
}

\usepackage{color}

\newcommand{\notE}{ \hbox{{$E$}\kern-.60em\hbox{/}}}
\newcommand{\notp}{\ \hbox{{$p$}\kern-.43em\hbox{/}}}

\newcommand{\eps}{\varepsilon}

\newcommand{\hp}{H^+}

\newcommand{\rtc}{\rho_{tc}}

\newcommand{\rtt}{\rho_{tt}}
\newcommand{\bbar}{\bar{b}}
\newcommand{\cbar}{\bar{c}}

\newcommand{\unit}[1]{\mathrm{#1}}

\newcommand{\tothe}[1]{\times 10^{#1}}

\preprint{\font\fortssbx=cmssbx10 scaled \magstep2
\hbox to \hsize{
\hskip1.2in 
\hbox{\fortssbx The University of Oklahoma}
\hskip0.2in $\vcenter{
                     \hbox{\bf arXiv: [hep-ph]}
                     \hbox{\bf OU-HEP-251123}
                     \hbox{November 2025}}$ }
}

\begin{document}

\title{\vspace*{0.7in}
  Enhanced Charged Higgs Signal at the LHC}

\author{
Chenyu Fang$^{a}$\footnote{E-mail address: chenyu.fang@ou.edu},
Wei-Shu Hou$^{b}$\footnote{E-mail address: wshou@phys.ntu.edu.tw},             
Chung Kao$^{a}$\footnote{E-mail address: Chung.Kao@ou.edu}, 
Mohamed Krab$^{b}$\footnote{E-mail address: mkrab@hep1.phys.ntu.edu.tw}
}

\affiliation{
$^a$Homer L. Dodge Department of Physics and Astronomy,
University of Oklahoma, Norman, OK 73019, USA \\
$^b$Department of Physics, National Taiwan University,
Taipei 10617, Taiwan}

\date{\today}

\bigskip

\begin{abstract}
  
 We investigate the discovery prospects of a charged Higgs boson ($H^\pm$)
 at the Large Hadron Collider (LHC) via the process
 $cg\to bH^\pm \to bc\bar{b}$ within the framework of a general two Higgs
 doublet model (G2HDM).
 In most two Higgs doublet models, the $H^+ cb$ coupling ($g_{H^+cb}$)
 is usually suppressed by the CKM matrix element $V_{cb}$.
 In G2HDM, there are additional Yukawa couplings,
 the process $cg\to bH^\pm \to bc\bar{b}$ is enhanced
 by the coupling $g_{H^+cb} \simeq \rtc V_{tb}$  
 in both the production and decay of the charged Higgs boson.
 We study possible physics backgrounds and evaluate the discovery
 potential with realistic acceptance cuts and tagging efficiencies at
 collider energies of $\sqrt{s}=$13 and 14 TeV.
 We apply $b$-tagging and $c$-tagging and show that $m_{H^+}$ can be
 extracted by pairing the tagged $b$ and $c$-jets.
 Our analysis leads to promising results for the current LHC and
 expected high-luminosity LHC.

\end{abstract}

\maketitle

\newpage

\section{Introduction}

The discovery of the 125 GeV Higgs boson in 2012~\cite{
 ATLAS:2012yve,CMS:2012qbp} marked the completion of the
experimentally observed particle spectrum
predicted by the electroweak theory of the Standard Model
(SM)~\cite{GLASHOW1961579,Weinberg:1967tq,Salam1968EW}.

However, certain phenomena, such as the baryon asymmetry of the
Universe~\cite{Sakharov:1967dj} as well as the flavor anomalies of
quarks and leptons~\cite{
  HeavyFlavorAveragingGroupHFLAV:2024ctg,Capdevila:2023yhq}, 
indicate deviations from the SM and suggest the need for additional
sources of Charge-Parity (CP) violation and extended Yukawa couplings.
These discrepancies highlight the necessity of extending SM and have
made the search for physics beyond the Standard Model (BSM) a primary
objective of the LHC.

The general two Higgs doublet model (G2HDM) extends the SM by
introducing an additional $SU(2)$ scalar doublet.
Following electroweak symmetry breaking, G2HDM gives rise to five physical
Higgs bosons: two CP-even scalars [$H^0$ (heavier) and $h^0$ (lighter)],
one CP-odd pseudoscalar ($A^0$) and a pair of charged Higgs bosons
($H^\pm$). Two scalar doublets can introduce sources of CP
violation~\cite{PhysRevD.42.860,FUYUTO2018402} and allow for a second
set of Yukawa couplings, including extra flavor-changing couplings.

To study the interaction of $\hp\bar{c}b$, we follow the Yukawa
Lagrangian present in Refs.~\cite{Davidson:2005cw,Mahmoudi:2009zx},
\begin{equation}\label{eq:yukawaL}
  \begin{aligned}
{\cal L}_Y =& \frac{-1}{\sqrt{2}} \sum_{\scalebox{0.6}{F=U,D,L}}
 \bar{F}\Big\{  \left[ \kappa^Fs_{\gamma}+\rho^F c_{\gamma} \right] h^0 +
   \left[ \kappa^Fc_{\gamma}-\rho^Fs_{\gamma} \right] H^0 \\
   &- i \, {\rm sgn}(Q_F)\rho^F A^0 \Big\} P_R F 
  -\bar{U} \left[ V \rho^D P_R - \rho^{U\dagger} V P_L \right] D H^+
  -\bar{\nu} \left[ \rho^L P_R \right] L H^+ + {\rm H.c.}, 
\end{aligned}
\end{equation}
where $P_{L,R} \equiv ( 1\mp \gamma_5 )/2$,
$c_{\gamma} \equiv \cos\gamma$,
$s_{\gamma} \equiv \sin\gamma$ with
$\gamma$ is the mixing angle between the two CP-even scalars $H$ and $h$,
and $Q_F$ is the fermion charge. 
The $\kappa^F$~matrices are diagonal and fixed by
fermion masses to $\kappa^F = \sqrt{2}m_F/v$ with $v \approx 246$ GeV, 
while the $\rho^F$~matrices contain both diagonal and off-diagonal
elements that serve as free parameters.
In our collider study, we choose $\rho^F$ as real.
In what follows, we drop the superscript $F$.

\begin{figure}[b]
\centering
\begin{tikzpicture}[font=\scriptsize]
  \begin{feynman}
    \vertex (a) at (0,2.2) {\(c\)};
    \vertex (b) at (0,0) {\(g\)};
    \vertex (c) at (0.8,1.1);
    \vertex (d) at (2,1.1);
    \vertex (e) at (2.8,2.2) {\(b\)};
    \vertex (f) at (2.8,0) {\(\hp\)};
    
    \diagram* {
      (a) -- [myarrow] (c), 
      (b) -- [gluon] (c), 
      (c) -- [myarrow, edge label'=\(c\)] (d),  
      (d) -- [myarrow] (e), 
      (d) -- [scalar] (f), 
    };
  \end{feynman}
\end{tikzpicture}
\begin{tikzpicture}[font=\scriptsize]
  \begin{feynman}
    \vertex (a) at (0,2.2) {\(c\)};
    \vertex (b) at (0,0) {\(g\)};
    \vertex (c) at (1.2,0.4);
    \vertex (d) at (1.2,1.8);
    \vertex (e) at (2.4,2.2) {\(\hp\)};
    \vertex (f) at (2.4,0) {\(b\)};
    
    \diagram* {
      (a) -- [myarrow] (d), 
      (b) -- [gluon] (c), 
      (d) -- [myarrow, edge label'=\(\bar{b}\)] (c),  
      (c) -- [myarrow] (f), 
      (d) -- [scalar] (e), 
    };
  \end{feynman}
\end{tikzpicture}
\caption{Leading-order Feynman diagrams for $cg\to b\hp$.}
\label{fig:feynman1}
\end{figure}
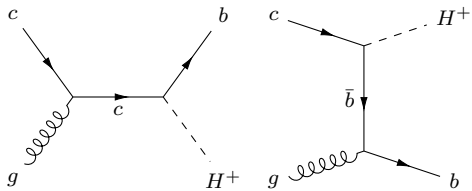

Focusing on the coupling of $H^+c\bar{b}$ one can derive the
interaction Lagrangian from Eq.~(\ref{eq:yukawaL}), 
\begin{equation}
\begin{aligned}
\mathcal{L}_{H^+ \bar{c}b} =&
-\,\bar{c} \left[ \left( V_{cd} \rho_{db} + V_{cs} \rho_{sb} + V_{cb} \rho_{bb} \right) P_R 
- \left( \rho^*_{uc} V_{ub} + \rho^*_{cc} V_{cb} + \rho^*_{tc} V_{tb} \right) P_L \right] b\, H^+ 
+ \text{H.c.} \\
\simeq& \;\bar{c} \left[ \rho^*_{tc} V_{tb} P_L \right] b\, H^+ 
+ \text{H.c.}.
\end{aligned}
\end{equation}
The coupling $\hp\bar{c}b \propto \rho_{tc} V_{tb}$ is not suppressed by CKM mixing~\cite{Ghosh:2019exx,Hou:2024bzh}. 
In this Letter, we propose the process $cg\to b\hp$ (see Fig.~\ref{fig:feynman1}) followed by $\hp \to c\bar{b}$ as a promising discovery channel of the BSM charged Higgs boson at LHC in the near future.
We first examine the parameter space of $\rtc$ under the current constraints imposed on the G2HDM. We then investigate the discovery potential of the Higgs signal in the presence of SM backgrounds. By applying realistic acceptance cuts, including $b$-tagging and $c$-tagging efficiencies, we present the statistical significance at the LHC for center-of-mass energies of $\sqrt{s} = 13$ and 14~TeV.
%

\section{Higgs Decays and Limits on Parameters}

In this section, we utilize the latest results from Higgs measurements at the LHC and $B$-physics experiments to constrain the parameter space of the G2HDM. 
We also present the branching ratios of the dominant decay channels of the charged Higgs with different parameters.

\subsection{Limits on $\rho_{tc}$}

Recent searches for the flavor-changing decay $t \to c h$ by ATLAS~\cite{ATLAS:2024mih}
and CMS~\cite{CMS:2024ubt} collaborations have placed stringent constraints on 
the branching ratio with ${\cal B}(t \to c h) \leq 3.4 \times 10^{-4}$~\cite{ATLAS:2024mih} and ${\cal B}(t \to c h) \leq 3.7 \times 10^{-4}$~\cite{CMS:2024ubt}, respectively.
This leads to an upper limit on the FCNH Yukawa coupling $|\lambda_{tch}|$
\begin{equation}
    \lambda_{tch} \leq 0.035
\end{equation}
for the effective Lagrangian,
\begin{equation}
  \mathcal{L} = -\frac{\lambda_{tch}}{\sqrt{2}}\bar{c}th^0 + \rm{H.c.},
  \label{eq:effLag}
\end{equation}
with the relation between $\lambda_{tch}$ and
the $t \to c h^0$ branching
ratio~\cite{ATLAS:2014lfm} being
\begin{equation}
    \lambda_{tch} \approx 1.92\times\sqrt{\mathcal{B}(t \to c h)},
\end{equation}
where $|\lambda_{tch}| = \tilde{\rho}_{tc}c_\gamma$ and
\begin{equation}
  \tilde{\rho}_{tc} \approx \sqrt{(|\rho_{tc}|^2 + |\rho_{ct}|^2)/2}.
\end{equation}
See Ref.~\cite{Krab:2025zuy} for the exclusion bounds in the $(c_\gamma, \rho_{tc})$ plane.

The effective Lagrangian for $H^0$ and $A^0$ can be written as
\begin{equation}
  \mathcal{L} = \frac{\lambda_{tcH}}{\sqrt{2}}\bar{c}t H^0
              +i\frac{\lambda_{tcA}}{\sqrt{2}}\bar{c}t A^0 + \rm{H.c.},
  \label{eq:effLag}
\end{equation}
where
\begin{equation}
  |\lambda_{tcH}| = \tilde{\rho}_{tc}s_\gamma
  \quad {\rm and} \quad 
  |\lambda_{tcA}| = \tilde{\rho}_{tc} \, .
\end{equation}
In the alignment limit~\cite{Craig:2013hca,Carena:2013ooa}, $c_\gamma \approx 0$ and $s_\gamma \approx 1$.
That leads to $|\lambda_{tch}| \to 0$, while $|\lambda_{tcH}|$ is
enhanced in the alignment limit, giving rise to same-sign top quark
and/or triple-top final states via
$cg \to tH^0 \to tt\bar c / tt\bar t$~\cite{Kohda:2017fkn,Hou:2020chc}.

ATLAS~\cite{ATLAS:2023tlp} and CMS~\cite{CMS:2023xpx} direct searches
for heavy Higgs bosons with flavor-changing couplings set further
limits on $\rho_{tc}$. Without interference between $H$ and $A$,
CMS~\cite{CMS:2023xpx} set no exclusion limits on $m_H$ or $m_A$ for a
coupling value $\rho_{tc} = 0.4$. When interference is included, still
assuming $\rho_{tc} = 0.4$, the limits reach $m_H = 290$~GeV and $m_A
= 340$~GeV. For a larger coupling, $\rho_{tc} = 1.0$, the limits
extend to $m_H = 760$~GeV and $m_A = 810$~GeV at 95\% confidence level
(CL)~\cite{CMS:2023xpx}.
ATLAS~\cite{ATLAS:2023tlp} excludes an $H$ with masses $m_H$ between
200-620~GeV at 95\% CL for coupling values $\rho_{tt} = 0.4$,
$\rho_{tc} = 0.2$, and $\rho_{tu} = 0.2$. For very small or vanishing
$\rho_{tu}$, these exclusion limits become
weaker~\cite{ATLAS:2023tlp}.

Flavor constraints on $\rho_{tc}$ are
weak~\cite{ALTUNKAYNAK2015135}. It is found that $B_s$ mixing
constrains $|\rho_{tc}| \lesssim 1.7$ for $m_{H^+} =
500$~GeV~\cite{Crivellin:2013wna,ALTUNKAYNAK2015135}. The coupling
$\rho_{ct}$ is constrained to be $\rho_{ct} <
0.1$~\cite{ALTUNKAYNAK2015135}.
In what follows, we set $\rho_{tc} = 0.4$, $\rho_{ct} = 0.05$. To be
consistent with flavor constraints, we choose $\rtt=0.2\times
(m_{H^+}/150~\unit{GeV})$~\cite{Hou:2019grj}.
The remaining Yukawa couplings, such as, e.g., $\rho_{bb}$ and
$\rho_{\tau\tau}$, are chosen to satisfy flavor constraints.
We refer to Ref.~\cite{Hou:2025tjp} for bounds on $\rho_{tt}$ from
$B$-physics and direct searches for $pp \to \bar t H^+ b \to \bar
tt\bar bb$~\cite{ATLAS:2021upq}.

\subsection{Charged Higgs decay channels}

\begin{figure}[t]
 \centering

 \includegraphics[height=60mm]{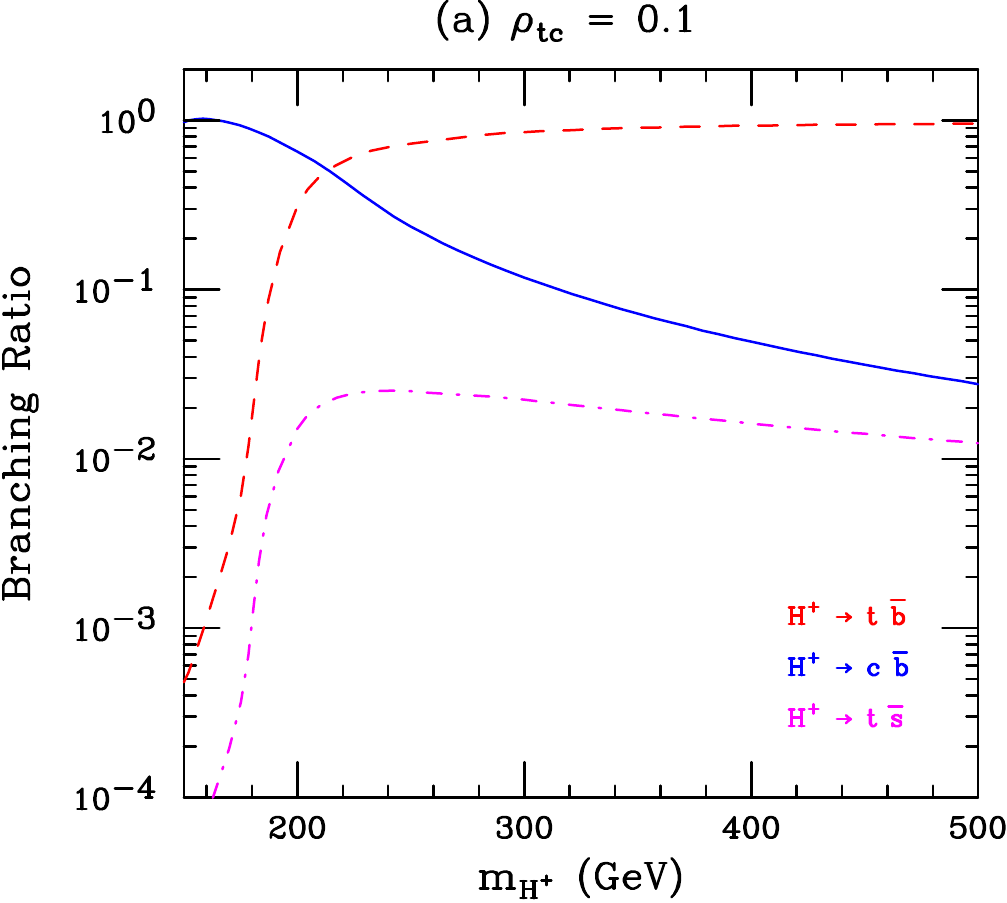}
 \includegraphics[height=60mm]{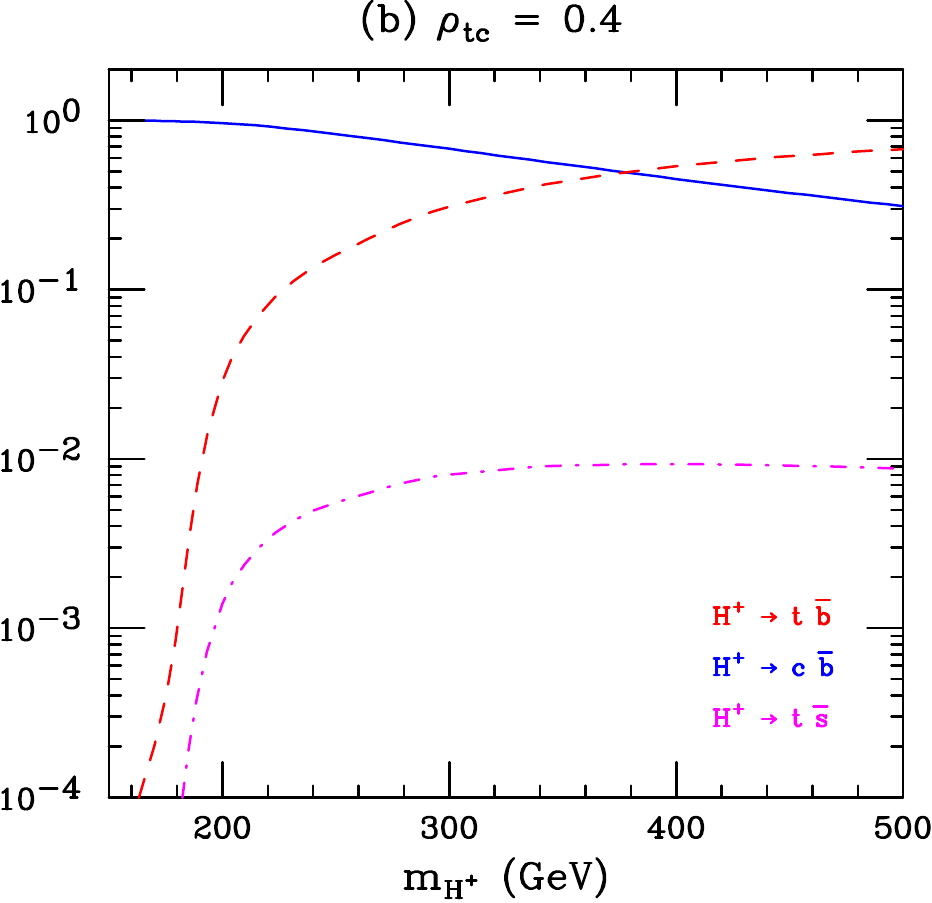}

 \caption{Branching ratios of $\hp$ as a function of $m_{\hp}$ for (a) $\rtc=0.1$, and (b) $\rtc=0.4$.}
   \label{fig:br_vs_rtc}
\end{figure}

We investigate various decay channels of the charged Higgs boson $\hp$, focusing specifically on $\hp \to c\bar{b}$, $\hp \to t\bar{b}$, and $\hp \to t\bar{s}$. For simplicity, we assume mass degeneracy among the additional Higgs bosons, i.e. $m_{H} \sim m_{A} \sim m_{\hp}$, to forbid $H^+ \to W^+ H, W^+ A$. However, if kinematically accessible, these decays could be promising discovery channels (see, e.g., Refs.~\cite{Bahl:2021str,Hou:2024ibt}).
The branching ratios are shown as functions of 
$m_{\hp}$ for fixed values of $\rtc = 0.1$ (left) and $\rtc = 0.4$ (right) in Fig.~\ref{fig:br_vs_rtc}. We see that while $\hp \to c\bar{b}$ becomes the dominant decay mode as $m_{\hp}$ less than 200 GeV for $\rtc = 0.1$ and less than 380~GeV for $\rtc\ge0.4$, the $\hp \to t\bar{b}$ channel remains competitive, especially for larger values of $m_{\hp}$.


\section{Higgs Signal and physics background}

\subsection{Higgs signal}
We study charged Higgs boson production at the LHC via the process $cg \to b\hp$, with the subsequent decay $H^+ \to c\bar{b}$, leading to a $bc\bar{b}$ final state. 
The complete Monte Carlo simulation is carried out using \textsc{MadGraph5\_aMC@NLO}~\cite{Alwall:2011uj,Alwall:2014hca}, \textsc{Pythia-8}~\cite{Sjostrand:2014zea}, and \textsc{Delphes}~\cite{deFavereau:2013fsa}.
The signal cross section is evaluated using the CT14LLO parton distribution function (PDF)~\cite{Dulat:2015mca}, with the renormalization and factorization scales set to $\mu_{R,F} = m_{H^+}$.
Applying the above scale choices and PDF, our
current estimates suggest a $K$-Factor of $\approx 1.95$, and is approximately
the same for all three energies ($\sqrt{s} = 13, 13.6,\text{and}\ 14$~TeV).
The $K$-factors are calculated using \textsc{MadGraph5\_aMC@NLO}.

\subsection{The Physics Background}

The dominant physics background to the signal arises from $ b\bbar j$ and $c\cbar j$.
Using the same simulation framework and PDF as for the signal, we compute the cross sections with both the renormalization and factorization scales set to $\mu_{R,F} = \sqrt{\hat{s}}$.
In addition, we scale our backgrounds to next-to-leading order using $K$-factor of 1.26 for $b\bbar j$ and $c\cbar j$~\cite{Alwall:2014hca}.

\subsection{Matching and Merging}

To ensure a consistent description between matrix-element calculations from \textsc{MG5}, which generate hard and well-separated particles, and the parton shower in \textsc{Pythia-8}, which models hadronization of quarks and glouns including soft or collinear emissions. Appropriate matching and merging schemes are employed, such as the MLM and CKKW-L algorithms~\cite{Mangano:2006rw, Lonnblad:2001iq}. These procedures prevent double counting of QCD emissions while preserving the correct jet multiplicity and kinematic distributions. The combination of \textsc{MadGraph} and \textsc{Pythia-8} thus provides a complete simulation framework, yielding fully hadronized final states that realistically represent the physical signatures expected at the LHC. In our study, we adopt the MLM matching scheme with \texttt{xqcut} = 25~GeV for both signal and background samples.

\subsection{Realistic Acceptance Cuts}

Using the default ATLAS Delphes card, we incorporate charm-tagging efficiencies based on recent CMS results~\cite{CMS:2021scf}, with $\eps_c=0.22$, $\eps_{b\to c}=0.01$, and $\eps_{j\to c}=0.001$, where $j = u,\,d,\,s,\,g$.
As for $b$-jet~\cite{ATLAS:2019bwq}, the $b$-tagging efficiency is approximately 0.70 ($\eps_b=0.70$),
the probability that a $c$-jet is mistagged as a $b$-jet 
is approximately 0.14 ($\eps_{c\to b}=0.14$), while
the probability that a light jet ($u, d, s, g$) is mistagged
as a $b$-jet is 0.01 ($\eps_{j\to b}=0.01$).
For the event analysis, each event is required to contain at least three jets,
including exactly one identified $c$-jet and two $b$-tagged jets.

We adopt the following basic requirements:
\begin{itemize}
\item[(i)] exactly one $c$-jet with $p^c_T \geq 25$ GeV, $|\eta_c|<2.5$;
\item[(ii)] exactly two $b$-jets with $p^b_T \geq 25$ GeV, $|\eta_b|<2.5$;
\item[(iii)] angular separation between every two pairs of jets satisfy $\Delta R_{jj} \geq$ 0.4.
\end{itemize}

\begin{figure}[t]
 \centering

 \includegraphics[width=67mm,height=60mm]{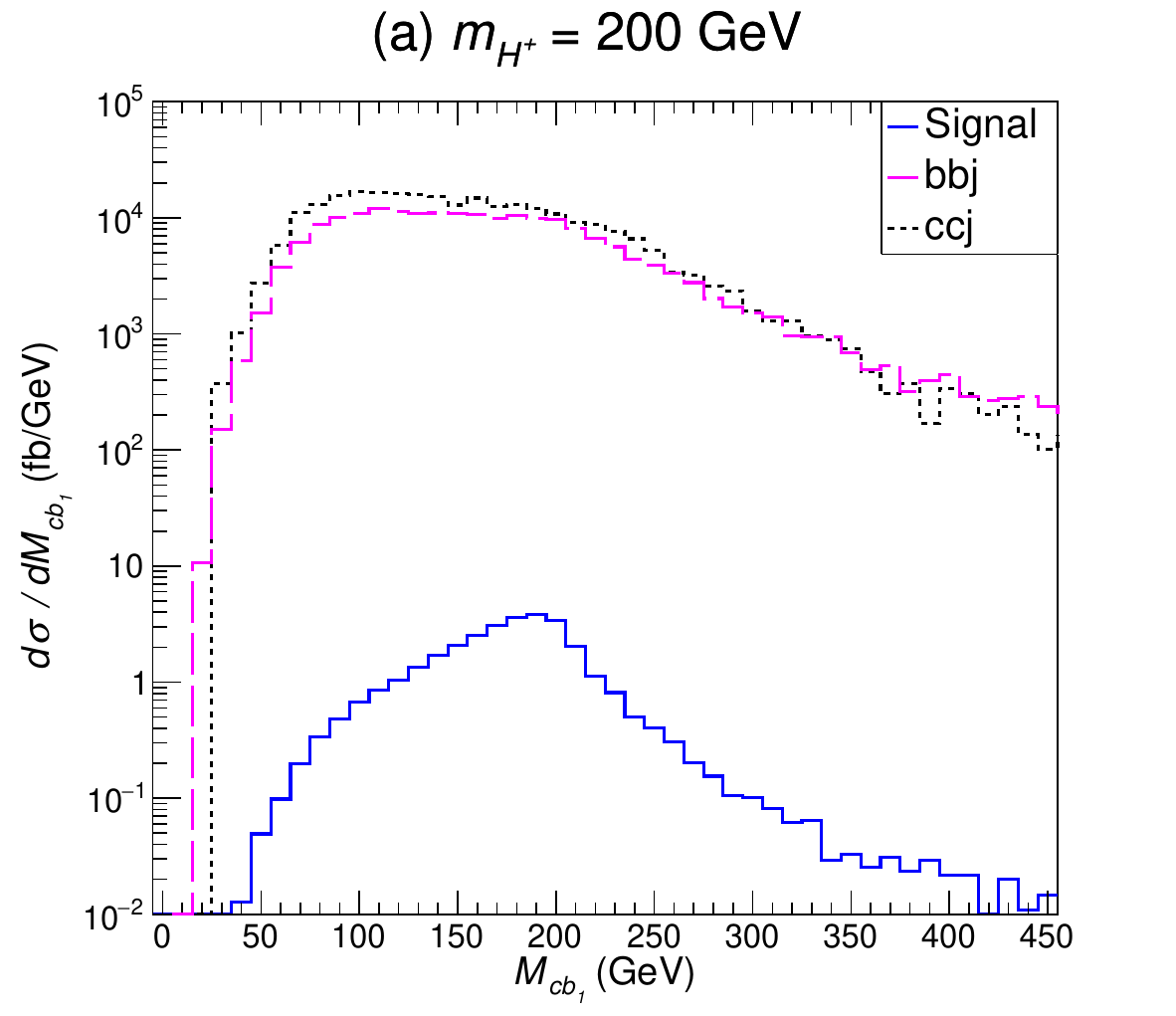}
 \includegraphics[width=65mm,height=60mm]{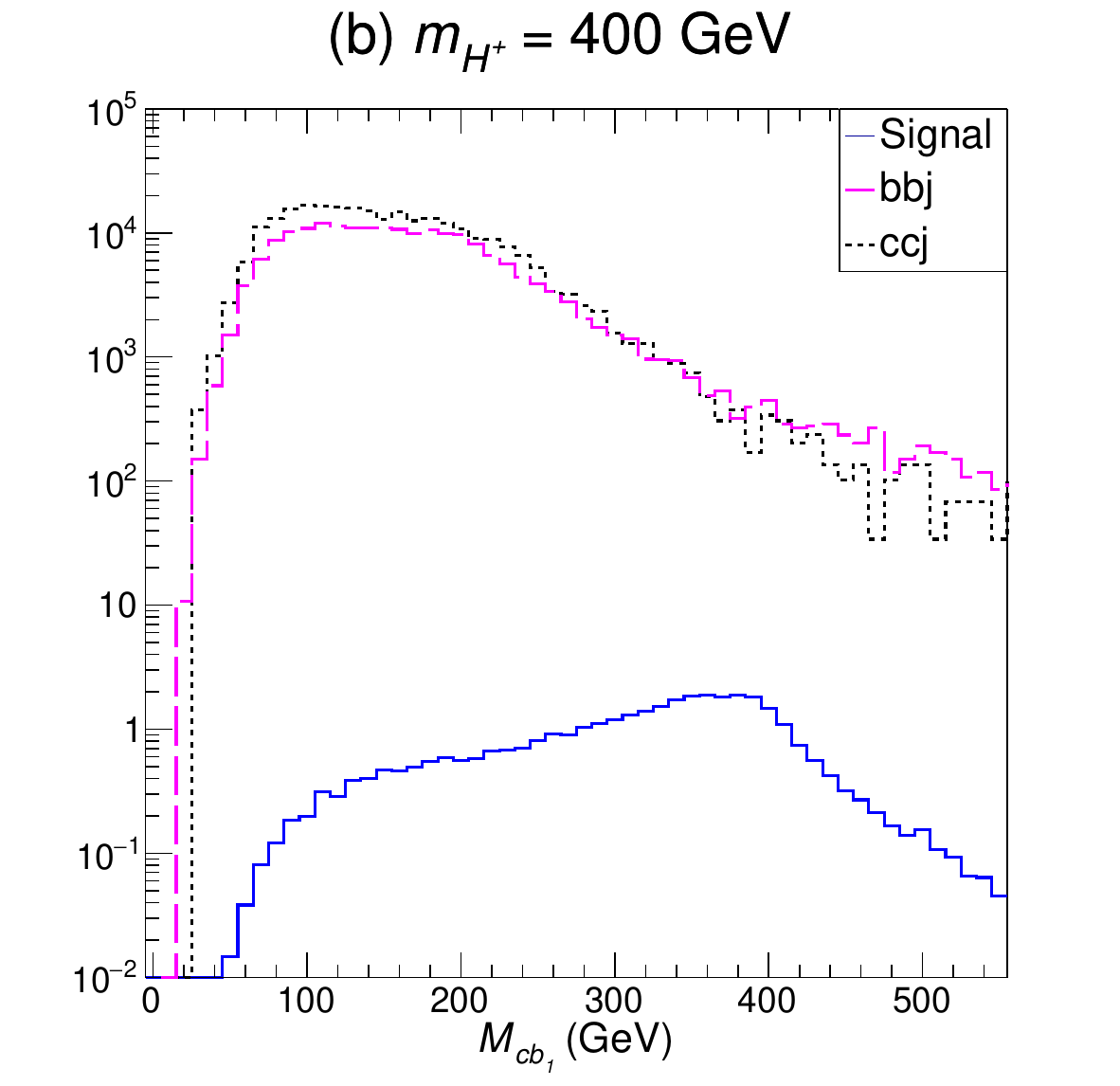}

 \caption{Invariant mass distributions $d\sigma/dM_{cb_1}$ for the signal and dominant backgrounds $b\bar{b}j$ (magenta, dashed) and $c\bar{c}j$ (black, dotted) at $\sqrt{s}=14$~TeV, shown for $\rho_{tc}=0.4$, (a) $m_{H^+}=200$~GeV (blue, solid) and (b) $m_{H^+}=400$~GeV (blue, solid). }
   \label{fig:mbc}
\end{figure}

To correctly reconstruct the events, 
the $c$-jet and one of the $b$-jets, denoted $b_1$, must originate from the decay of the charged Higgs boson ($H^+$) and are expected to be energetic.
In addition to the basic selection cuts, 
we further require $p^{b_1}_T > 50$ GeV, $p^c_T > 70$ GeV and an angular separation $\Delta R_{cb_1}>2.0$.
The invariant mass distribution of the $c$-jet and $b_1$-jet pair, $M_{cb_1}$, are expected to exhibit a peak around the mass of the charged Higgs, $m_{H^+}$. Fig.~\ref{fig:mbc} shows the $d\sigma/M_{cb_1}$ distributions for the signal with $m_{H^+}=200$ and $400$~GeV, together with the dominant physics backgrounds. 
A pronounced peak appears near $m_{H^+}$ for $m_{H^+}= 200$~GeV case, while for 400~GeV the peak is less distinct, but still provides discrimination against backgrounds.
Using the ATLAS mass resolution~\cite{ATLAS:2020bhl},
the reconstructed $c$-jet and $b_1$-jet masses are required to lie a 
mass window of $\Delta M_{cb_1} = 0.20\times m_{\hp}$. 
These selection requirements are optimized to suppress background processes while enhancing the statistical significance of the signal. The complete set of final selection criteria is summarized in Table~\ref{allcuts}.

%
%
\begin{table}[t]
 \centering
 \begin{tabular}{c} \hline 
 selection rule  \\
   \hline 
 1 tagged $c$-jet, 2 tagged $b$-jets \\
  $p^j_T>25$ GeV, $|\eta_j|<2.5$, $\Delta R_{jj}>$0.4\\
  $p^{b_1}_T>50$ GeV \\
  $p^c_T>70$ GeV \\
  $\Delta R_{cb_1}>2.0$ \\
  $|M_{c b_1} - m_{\hp}| \leq 0.20 \times m_{\hp}$ \\
 \hline 
 \end{tabular}
 \caption{Final selection criteria for signal and backgrounds events. }
 \label{allcuts}
\end{table}

\section{Discovery Potential}

%
%
\begin{table}[t]
  \centering
  \setlength{\tabcolsep}{8pt}
  \begin{tabular}{cccccc}
  \hline\noalign{\vskip .5mm}
  $\sqrt{s}$ (TeV)  & $\text{BP}_1$ (fb)  
     & $b \bar{b}j$ (fb)  & $c\bar{c}j$ (fb) & Total (fb) & $N_{SS}$ \\
    \hline
  13 & 353.4 & 8.052$\tothe{4}$  & 7343
     & 8.786$\tothe{4}$ & 26.6 \\
  13.6 & 394.0  & 8.407$\tothe{4}$ & 7851 
     & 9.192$\tothe{4}$  & 28.9 \\
  14 & 406.3 & 8.988$\tothe{4}$ & 8741
     & 9.862$\tothe{4}$   & 28.9 \\
 \hline \noalign{\vskip .5mm}
  $\sqrt{s}$ (TeV)  & $\text{BP}_2$ (fb)  
     & $b \bar{b}j$ (fb) & $c\bar{c}j$ (fb)  & Total (fb) & $N_{SS}$ \\
    \hline
  13 & 103.2  & 8.332$\tothe{4}$ & 7692
     & 9.102$\tothe{4}$ & 7.65 \\
  13.6 & 109.4  & 9.009$\tothe{4}$ & 8223 
     & 9.831$\tothe{4}$ & 7.80 \\
  14 & 116.1 & 9.723$\tothe{4}$ & 9178
     & 1.064$\tothe{5}$  & 7.96 \\
 \hline\noalign{\vskip .5mm}
  $\sqrt{s}$ (TeV)  & $\text{BP}_3$ (fb) 
     & $b \bar{b}j$ (fb)  & $c\bar{c}j$ (fb) & Total (fb) & $N_{SS}$ \\
    \hline
  13 & 25.61 & 6.323$\tothe{4}$ & 5634
     & 6.887$\tothe{4}$ & 2.18 \\
  13.6 & 27.64 & 6.927$\tothe{4}$ & 5762
     & 7.503$\tothe{4}$ & 2.26 \\
  14 & 29.37 & 7.129$\tothe{4}$ & 7025
     & 7.832$\tothe{4}$ & 2.35\\
 \hline
  \end{tabular}
  \caption{Cross sections for signal and backgrounds for $\rho_{ct}=0.05$, $\rho_{tc}=0.4$, $m_{\hp} = 200$~GeV ($\rm BP_1$), $m_{\hp} = 300$~GeV ($\rm BP_2$), and $m_{\hp} = 400$~GeV ($\rm BP_3$). Statistical significance is presented with integrated luminosity $L=500\, \text{fb}^{-1}$. $K$-factors are included in the cross section.}
  \label{sigcross}
\end{table}

After applying all selection criteria at collider energies $\sqrt{s} = 13$, 13.6, and 14~TeV, the signal cross sections for various benchmark points (BPs), along with the corresponding background cross sections, are summarized in Table~\ref{sigcross}.
The estimated statistical significance ($N_{SS}$), calculated following Ref.~\cite{Cowan:2010js},
\begin{equation}
    N_{SS} = \sqrt{2 (N_S + N_B)\ln(1 + N_S/N_B) - 2  N_S}
\end{equation}  
are also presented in Table~\ref{sigcross},
where $N_{S}$ and $N_{B}$ are number of signal and background events, respectively.


\begin{figure}[htb]
    \centering
    \includegraphics[width=80mm,height=66mm]{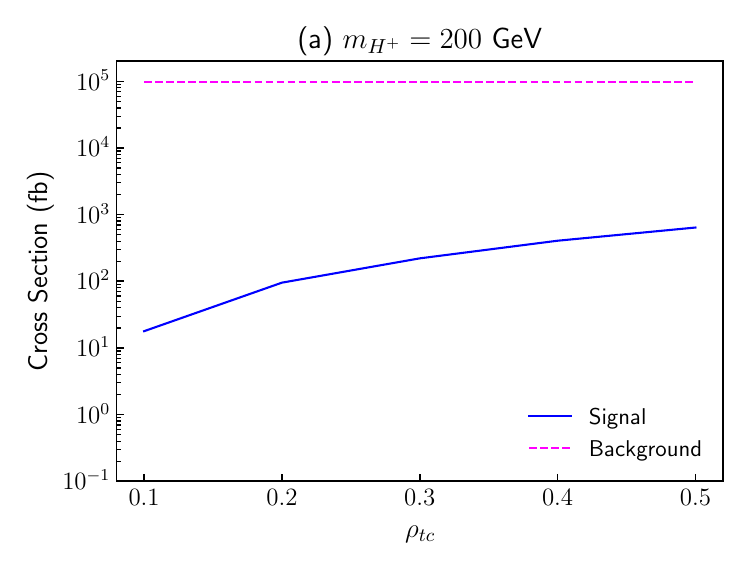}
    \includegraphics[width=76mm,height=66mm]{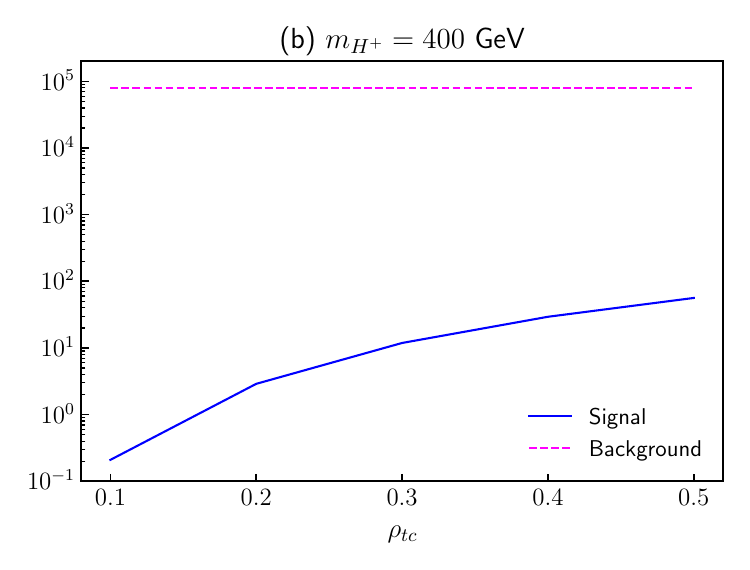} 
    \caption{The cross section as a function of $\rho_{tc}$ for signal and background at $\sqrt{s} =$ 14 TeV after selection cuts, shown for (a) $m_{\hp}$ = 200~GeV and (b) $m_{\hp}$ = 400~GeV.}
    \label{fig:svr}
\end{figure}


\begin{figure}[t]
    \centering
    \includegraphics[width=80mm,height=66mm]{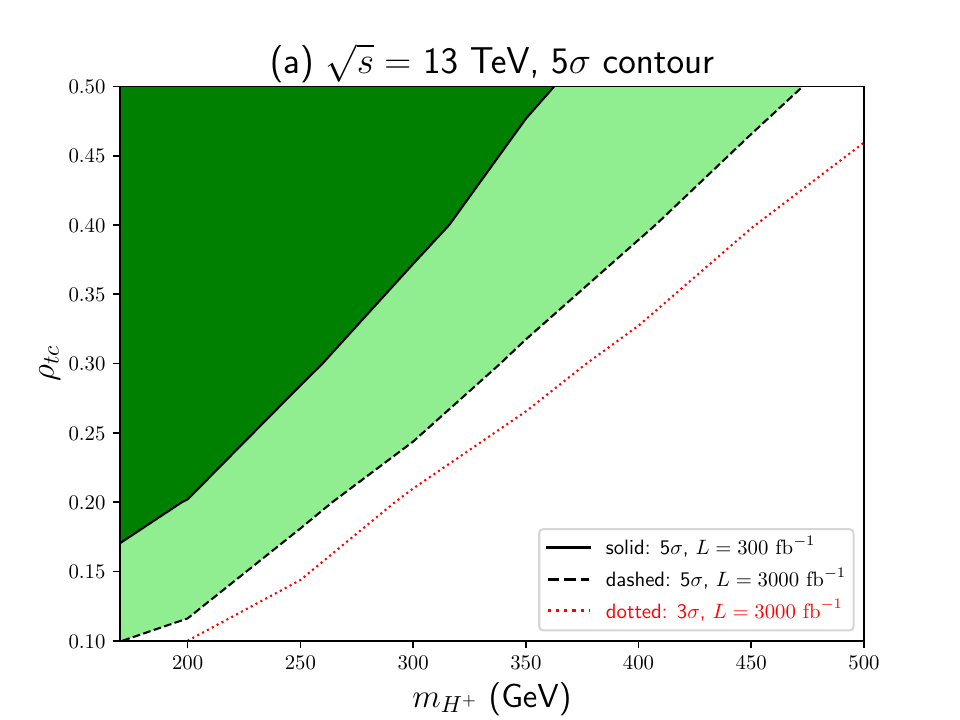}
    \includegraphics[width=80mm,height=66mm]{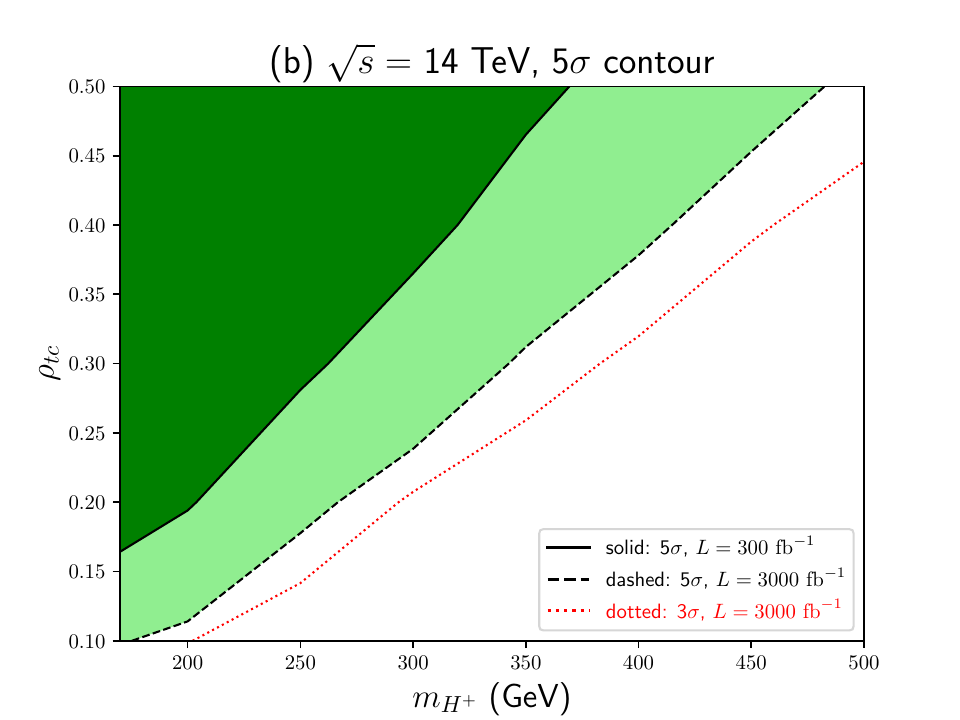} 
    \caption{The 5$\sigma$ discovery contours at the LHC
      in the $(m_{\hp},\rho_{tc})$ plane
      for (a) $\sqrt{s} =$ 13 TeV and (b) $\sqrt{s} =$ 14 TeV 
      with $L = 300\, \text{fb}^{-1}$ (dark green solid) and
      $L = 3000\, \text{fb}^{-1}$ (light green dashed). 
      In addition, 3$\sigma$ discovery contour with $L = 3000\, \text{fb}^{-1}$ (red dotted) are shown for both 13 and 14 TeV.}
    \label{fig:contours}
\end{figure}

Fig.~\ref{fig:svr} presents the cross section as a function of $\rho_{tc}$ for different values of $m_{\hp}$ at a collider energy of $\sqrt{s}=$ 14~TeV.
Fig.~\ref{fig:contours} presents the $5\sigma$ discovery contour at
the LHC for (a) $\sqrt{s} =$ 13 TeV and (b) $\sqrt{s} =$ 14 TeV
in the $(m_{\hp},\rho_{tc})$ plane,
assuming integrated luminosities of $L=300$ and $3000$~fb$^{-1}$.
The results clearly demonstrate that 14~TeV LHC with high luminosity, $L=3000$~fb$^{-1}$, substantially enhances the discovery potential.

\section{Conclusions}

In the G2HDM, the process $H^+ \to c\bar{b}$ provides an excellent channel for probing the charged Higgs boson.
The coupling $\hp \cbar b$ depends strongly on $\rho_{tc}V_{tb}$, making this process particularly sensitive to place constraints on $\rho_{tc}$ as well.
For a heavy charged Higgs boson with $m_{H^+} > m_t$, the $b$- and $c$-jets from its decay are energetic, making the signal more distinguishable from the physics backgrounds.
When combined with advanced tagging techniques, the search for $H^+$ becomes especially promising.

We have investigated the discovery prospects at the LHC, focusing on the process $pp \to bH^+ \to bc\bar{b}$.
The primary physics background comes from $b\bbar j$ and $c\cbar j$. 
Both signal and background are analyzed under realistic selection criteria at collider energies of $\sqrt{s}=13$, 13.6, and 14~TeV.
According to existing constraints on the model parameters from recent studies, we evaluate the cross sections and statistical significance for various BPs,
and present the corresponding discovery contours in the $(m_{H^+}, \rho_{tc})$ plane.
Our analysis shows that, at $\sqrt{s}=14$~TeV, the signal remains observable for $\rho_{tc}\gtrsim 0.1$ and $m_{H^+}\lesssim450$~GeV with an integrated luminosity of $L=3000~\text{fb}^{-1}$.

In summary, we have made several important contributions to the search for the charged Higgs boson:
\begin{itemize}
\item[(i)] We identify a promising production channel that is not
  suppressed by the CKM matrix, in contrast to two Higgs doublet models
  with the so-called natural flavor conservation, such as the popular 2HDM
  type-II.
\item[(ii)] The signal cross section depends solely and strongly on
  $\rho_{tc}$, thereby providing a powerful method to further test the
  G2HDM and constrain this parameter.
\item[(iii)] We show that $m_{H^+}$ can be extracted by pairing the
  tagged $b$ and $c$-jets, which provides discrimination against
  physics backgrounds.

\end{itemize}

\section*{Acknowledgments}

CK would like to thank George Hou and the High Energy Physics Group at
National Taiwan University for excellent hospitality, where part of
the research was completed.
This research is supported in part by the University of Oklahoma (FC and CK),
the National Science and Technology Council of Taiwan under grant
No.~114-2639-M-002-006-ASP (WSH and MK), and NTU grants No.~114L86001
and No.~114L891801 (WSH).


\bibliographystyle{JHEP}  
\bibliography{references}  

\end{document}